\documentclass[twocolumn,showpacs,preprintnumbers,amsmath,amssymb,prb,floatfix]{revtex4-1}

\usepackage{amssymb}
\usepackage{graphicx}
\usepackage{bm}% bold math

\begin{document}
\title{Effective carrier type and field-dependence of the reduced-T$_c$ superconducting state in SrFe$_{2-x}$Ni$_{x}$As$_{2}$}

\author{N. P. Butch}
\email{nbutch@umd.edu}
\author{S. R. Saha}
\author{X. H. Zhang}
\author{K. Kirshenbaum}
\author{R. L. Greene}
\author{J. Paglione}
\affiliation{Center for Nanophysics and Advanced Materials, Department of Physics, University of Maryland, College Park, MD 20742}
\date{\today}

\begin{abstract}
Measurements of the Hall effect, thermoelectric power, magnetic susceptibility and upper and lower critical fields were performed on single crystals of SrFe$_{2-x}$Ni$_{x}$As$_{2}$, an FeAs-based superconducting system that exhibits a reduced superconducting transition temperature $T_c$ in comparison to most other iron-pnictide superconductors. Studies of the Hall and thermoelectric responses indicate that Ni substitution in this system results in a dominant electron-like response, consistent with electron doping in other similar systems but with a weaker change in the Hall coefficient and a more gradual change in the thermoelectric response with Ni concentration. For optimally doped samples with full superconducting volume fraction, the lower and upper critical fields were determined to be $H_{c1}(1.8$~K$)= 0.08$~T and  H$_{c2}(0)= 25$~T, respectively, with lower-$T_c$ samples showing reduced values and indications of inhomogeneous superconductivity.  Comparable to other higher-$T_c$ FeAs-based materials, the temperature dependence of the upper critical field, $\partial{H_{c2}}/\partial{T}$, is linear over a wide temperature range, and the large values of H$_{c2}(0)$ greatly exceed conventional estimates of paramagnetic and orbital limits.

\end{abstract}

\pacs{74.70.Xa,74.25.F-,74.25.Ha,72.15.Jf}
\maketitle

\section{Introduction}
Chemical substitution in the iron pnictide ``122'' compounds, which form in the tetragonal ThCr$_2$Si$_2$ crystal structure, has uncovered a large class of superconductors. The superconducting state in these compounds is a topic of great interest, as it appears to have unconventional pairing symmetry\cite{Parker09,Chubukov09} and exists in proximity to magnetic order.  Understanding superconductivity in the iron pnictides may also offer insight into the unusual superconducting states in other classes of materials, such as the high-$T_c$ cuprates, heavy fermion intermetallics, and organic charge-transfer salts.\cite{Butch08} Transition metal substitution can effectively electron-dope the antiferromagnetic parent compound, yielding superconductivity when Co, Ni, Ru, Rh, Pd, and Ir replace Fe.\cite{Saha09,Han09,Schnelle09,Ni09}  In contrast, significantly higher transition temperatures of 35~K are achieved via hole-doping on the alkaline-earth site, although these values are still lower than the 55~K superconducting transitions seen in the related ``1111'' materials.

Despite their lower values of superconducting transition temperature $T_c$ relative to the 1111 materials, the 122 compounds can be prepared as large single crystals and are well-suited for experimental study.  One particularly interesting aspect of the superconductivity in 122 materials is the similarity of maximal $T_c$, 20-25~K, regardless of the transition metal substituent.  In fact, this trend is known to be broken only in the case of SrFe$_{2-x}$Ni$_{x}$As$_{2}$~\cite{Saha09} and SrFe$_{2-x}$Pd$_{x}$As$_{2}$,\cite{Han09} both exhibiting $T_{c} < 10$~K.

\begin{figure}%[hp]
    {\includegraphics[width=3.4in]{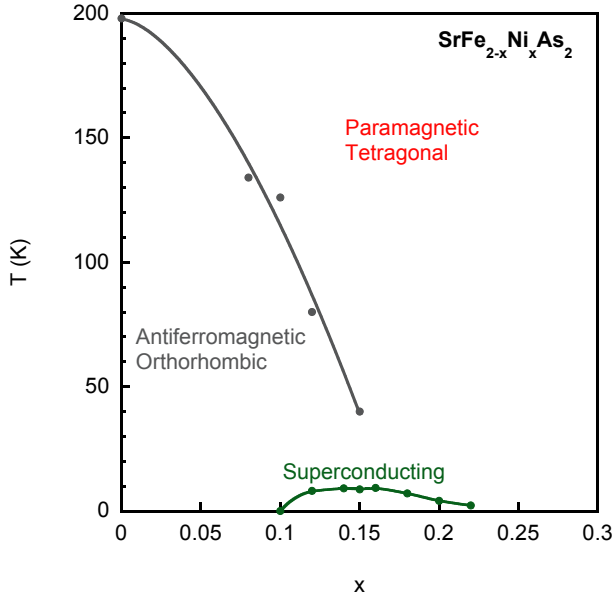}}
    \caption{(Color online) The phase diagram of SrFe$_{2-x}$Ni$_{x}$As$_{2}$ as determined in Ref.~\onlinecite{Saha09}. The magneto-structural transition is not observed for $x>0.15$ and superconductivity is found for $0.10 \leq x \leq 0.22$. The maximum $T_c \approx 9$~K.}
    \label{phsdgm}
\end{figure}

The temperature-chemical concentration phase diagram of SrFe$_{2-x}$Ni$_{x}$As$_{2}$ (Fig.~\ref{phsdgm}) was recently studied.\cite{Saha09} As with other iron pnictide superconductors, Ni substitution into SrFe$_{2}$As$_{2}$ initially suppresses the magneto-structural transition temperature $T_0$, which for SrFe$_{2}$As$_{2}$ occurs at about 200~K. The transition can be tracked to $x=0.15$, where $T_0 \approx 40$~K, but it is not observed in $x=0.16$.   Superconductivity is observed in the concentration range $0.10 \leq x \leq 0.22$, with the highest values of superconducting transition temperature $T_c \approx 9$~K for $0.15 \leq x \leq 0.18$.  Bulk superconductivity is confirmed at $x=0.15$ by the presence of full diamagnetic screening and a small specific heat anomaly at $T_c$, whose magnitude is consistent with those observed in other 122 materials with comparable $T_c$ values.\cite{Budko09} The superconducting ``dome'' is asymmetric, with rather sharp onset and more gradual offset of superconductivity as a function of $x$.  At the edges of the dome, the width in $T$ of the superconducting transitions increases, and the diamagnetic screening fraction is substantially decreased.  These characteristics may be taken as signatures of inhomogeneous superconductivity appearing at the edges of the superconducting phase region, even though the Ni distribution appears chemically homogeneous throughout the entire substitutional range.

In order to understand whether the reduced values of $T_c$ in SrFe$_{2-x}$Ni$_{x}$As$_{2}$ are entirely coincidental, or whether there is something fundamentally different about the superconductivity in this system, a study of the effective carrier type and the properties of the superconducting state was carried out on single crystals of SrFe$_{2-x}$Ni$_{x}$As$_{2}$.

\section{Experimental methods}
Single crystals of SrFe$_{2-x}$Ni$_{x}$As$_{2}$ were synthesized in excess FeAs and annealed in inert atmosphere.  Details of sample preparation and characterization have been published.\cite{Saha09} Actual Ni concentration $x$ was found to closely match nominal concentration. Measurements of the electrical resistivity, Hall effect, and thermoelectric power were carried out in a 14~T Quantum Design Physical Property Measurement System.  Electrical resistivity was measured via a 4-probe technique using low-frequency ac currents of 100~$\mu$A.  Thermoelectric power was measured by applying a constant temperature difference of 0.7~K across each sample. Magnetization measurements were performed in a 7~T Quantum Design Magnetic Property Measurement System.

\begin{figure}%[hp]
    {\includegraphics[width=3.4in]{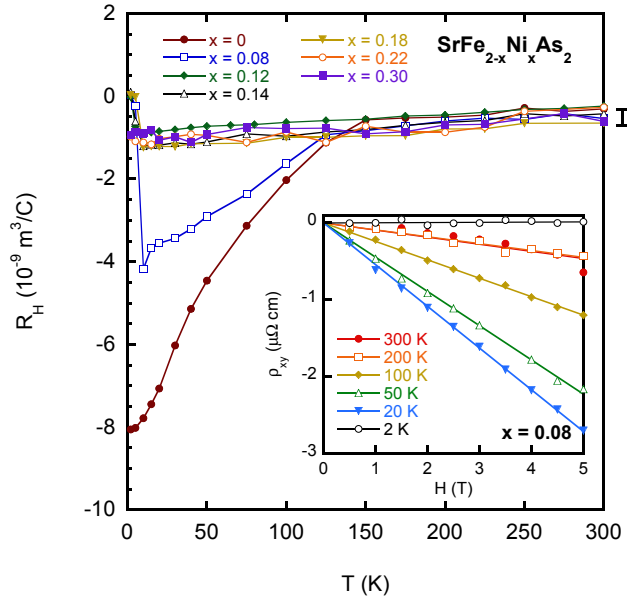}}
    \caption{(Color online) Hall coefficient of SrFe$_{2-x}$Ni$_{x}$As$_{2}$ as a function of temperature. The magnetic transitions in $x=0$ and $x=0.08$ are evident as a drop in $R_H$. Superconducting transitions are evident for $0.08 \leq x \leq 0.22$ at low temperatures. All measurements are consistent with a dominant contribution from negative charge carriers. The error bar on the right illustrates the window of uncertainty of about $\pm 0.2 \times 10^{-9}$~m$^{3}$/C. Inset: Field dependence of the transverse resistivity is linear to at least 5~T.}
    \label{Hall}
\end{figure}

\section{Hall effect and thermoelectric power}
A comparison of the temperature $T$ dependence of the Hall constant $R_H$ for SrFe$_{2-x}$Ni$_{x}$As$_{2}$ is shown in Fig.~\ref{Hall}. The value of $R_H$ was calculated from the slope of the symmetrized transverse electrical resistivity $\rho_{xy}$ collected in magnetic fields $-5$~T$ \leq H \leq 5$~T.  As shown in the inset of Fig.~\ref{Hall}, the $\rho_{xy}(H)$ data are linear with a negative slope in $H < 5$~T, indicating the existence of a dominant electron-like signal.  For all Ni concentrations, these values of $R_H$ at room temperature correspond to a density of carriers of about $10^{22}$~cm$^{-3}$ in a one-band model.  For $x=0$ and $x=0.08$, a change in carrier concentration coincides with the magneto-stuctural transition, yielding a low-$T$ carrier concentration of $10^{21}$~cm$^{-3}$.  For higher $x$, $R_H$ is remarkably $T$-independent, although the superconducting transition is readily discernable. A clear trend towards lower or higher carrier number with Ni substitution can not be identified in these data. The uncertainty in the measurements is estimated at $\pm 0.2 \times 10^{-9}$~m$^{3}$/C, potentially masking any real $x$-dependence in the data. In contrast, in the case of BaFe$_{2-x}$Co$_{x}$As$_{2}$, such a trend is evident, with the variation in $R_H \geq 1 \times 10^{-9}$~m$^{3}$/C as a function of Co concentration.\cite{Rullier-Albenque09,Fang09} While more precise Hall measurements are required to reach a conclusion about the actual variation in $R_H$ for SrFe$_{2-x}$Ni$_{x}$As$_{2}$, it seems clear that the magnitude is significantly smaller than that observed for the BaFe$_{2-x}$Co$_{x}$As$_{2}$ system. For SrFe$_2$As$_2$, the low-$T$ value of $R_H$ determined in this study is slightly greater than half of a previously published value, $-13\times10^{-9}$~m$^3$/C,\cite{Yan08} and significantly less than $-25\times10^{-9}$~m$^3$/C found for BaFe$_2$As$_2$,\cite{Rullier-Albenque09,Fang09,Mun09} although the complex Fermi surface of these materials makes it difficult to make any direct comparisons of carrier density between Ba and Sr materials.  These results are also consistent with ARPES results showing that there is a dominant electron Fermi surface in Co-substituted BaFe$_2$As$_2$.\cite{Brouet09,Liu09}

\begin{figure}%[hp]
    {\includegraphics[width=3.4in]{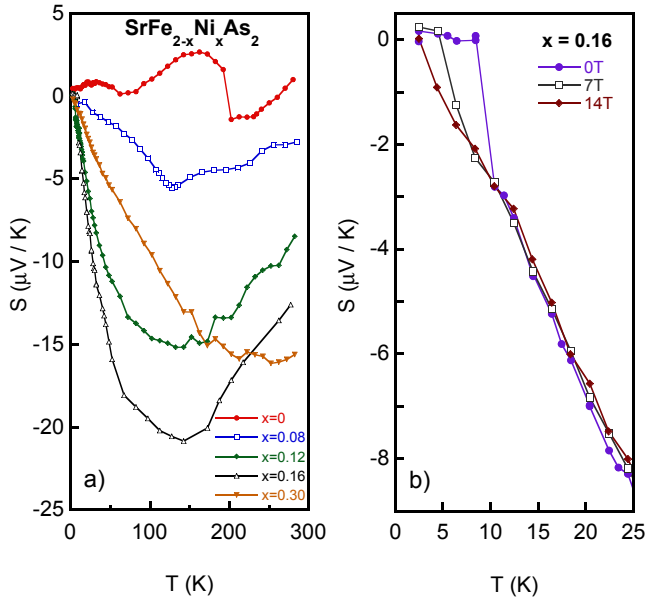}}
    \caption{(Color online) a) Thermoelectric power $S$ of SrFe$_{2-x}$Ni$_{x}$As$_{2}$ vs temperature. The Ni-substituted samples exhibit linear $T$-dependence at low temperatures, as would be expected in an electron-dominated metal.  Slopes increase with increasing Ni content for $x<0.16$. b) Magnetic field dependence of $S(T)$ illustrates the suppression of superconductivity for $x=0.16$.}
    \label{TEP}
\end{figure}

The temperature dependence of the thermopower $S$ is shown in Fig.~\ref{TEP}a for several values of $x$.  The data qualitatively resemble those of Ba(Fe$_{1-x}$Co$_x$)$_2$As$_2$, which is superconducting, and Ba(Fe$_{1-x}$Cu$_x$)$_2$As$_2$, in which no trace of superconductivity has been found.\cite{Mun09}  For the parent compounds CaFe$_2$As$_2$, SrFe$_2$As$_2$ and BaFe$_2$As$_2$, the magneto-structural transition is evident in $S(T)$, and at lower temperatures, the $T$ dependence is non-monotonic. In SrFe$_2$As$_2$, $S(T)$ changes sign, which is also observed in CaFe$_2$As$_2$.\cite{Matusiak09}  Upon electron doping, regardless of the alkaline earth species, the thermopower is negative below room temperature.  In SrFe$_{2-x}$Ni$_{x}$As$_{2}$, the low $T$ thermopower data increase in magnitude  monotonically, although a change of slope occurs at a higher $x$-dependent temperature.  In the case of $x=0.08$, the local extremum in $S(T)$ corresponds to the magneto-structural transition.  With increasing $x$, the extremum occurs at increasing $T$. However, at higher $x$ it is not obviously correlated with any features in transport, magnetization, or heat capacity.  While it is possible that the $S(T)$ minima might arise from changes in the Fermi surface, there are no corresponding features in the Hall data. An alternative explanation is that the relative mobilities of the various carrier species are changing.  The magnitude of $S(T)$ is maximal with a value of about -20~$\mu$V/K at $x=0.16$, which has the highest superconducting transition temperature $T_c$. This value of $S$ is less than half that of the maximum value in Ba(Fe$_{1-x}$Co$_x$)$_2$As$_2$, and is smaller also than that of Ba(Fe$_{1-x}$Cu$_x$)$_2$As$_2$.  As in the Ba122 case, both $R_H(T)$ and $S(T)$ data sets are difficult to explain because there are multiple Fermi surfaces and the $x$-dependence of magnetic interactions is complicated.  However, the $x$-dependent change in $S(T)$ noted in the Ba compounds is not readily apparent in SrFe$_{2-x}$Ni$_{x}$As$_{2}$, which features a more gradual change with $x$.

Fig.~\ref{TEP}b shows the magnetic field $H$ dependence of  $S(T)$ for $x=0.16$.  Here, $H$ was applied along the $c$-axis. The superconducting transition is clearly visible, as its suppression by applied field, which is consistent with the electrical resistivity studies presented next.

\begin{figure}%[hp]
    {\includegraphics[width=3.4in]{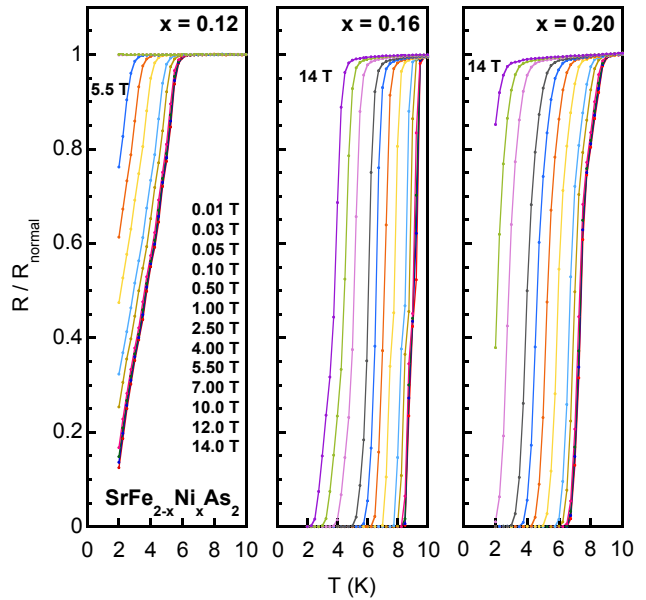}}
    \caption{(Color online) Electrical resistivity of SrFe$_{2-x}$Ni$_{x}$As$_{2}$ vs temperature, showing the suppression of the superconducting transition by increasing magnetic field, applied parallel to the $c$-axis. The resistivity is normalized to the normal state value just above the transition.  For $x=0.12$, 7~T data are not shown.}
    \label{Hc2sweeps}
\end{figure}

\section{Upper critical field}
The suppression with applied magnetic field $H$ of the resistive superconducting transition of SrFe$_{2-x}$Ni$_{x}$As$_{2}$ is illustrated in Fig.~\ref{Hc2sweeps} for under- ($x=0.12)$, optimally- ($x=0.16)$, and over-doped ($x=0.20)$ concentrations.  The data are normalized to the normal-state resistance for clarity.  There is a clear qualitative change in the superconducting transitions across the superconducting concentrations.  As noted previously,\cite{Saha09} the transition width is narrowest near optimal Ni concentrations.  In fact, for $x=0.12$, the resistive transition is incomplete down to 1.8~K, as shown in Fig.~\ref{Hc2sweeps}, which may be due to macroscopic phase separation between superconductivity and antiferromagnetism, as is observed in Ba$_{1-x}$K$_x$Fe$_2$As$_2$ by $\mu$SR measurements.\cite{Park09} Note also that this is consistent with the zero-field $R(T)$ transition presented earlier,\cite{Saha09} in which a second step is clearly visible. This possibility is an interesting contrast to the case of Ba(Fe$_{1-x}$Co$_x$)$_2$As$_2$, which instead supports the microscopic coexistence of superconducting and magnetic order.\cite{Pratt09,Laplace09} While local probes such as $\mu$SR or NMR are required for confirmation of true microscopic coexistence, these data suggest that electronic phase separation can arise also in the case of transition metal substitution.

\begin{figure}%[hp]
    {\includegraphics[width=3.4in]{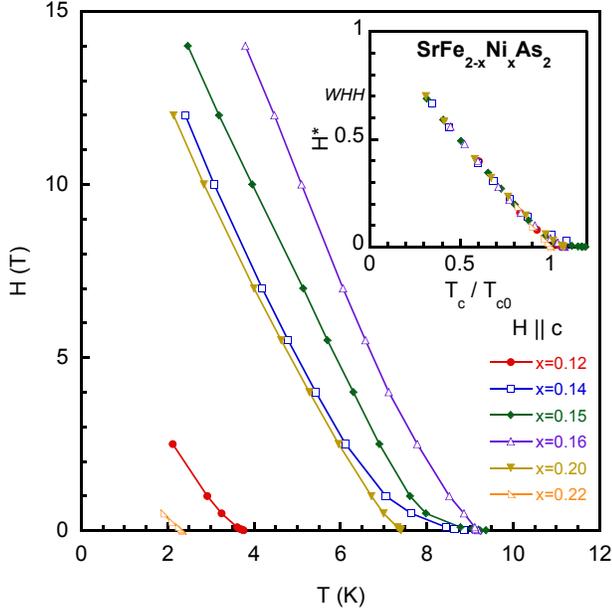}}
    \caption{(Color online) Upper critical field $H_{c2}$ of SrFe$_{2-x}$Ni$_{x}$As$_{2}$, for $H \parallel c$, vs temperature. The points here denote half of the resistive transition.  Inset:  $H_{c2}$ curves scaled to the zero field transition temperature $T_{c0}$ and a slope of -1.  The value of  $H_{c2}$ at $T=0$~K exceeds the WHH estimate.}
    \label{Hc2}
\end{figure}

\begin{figure}%[hp]
    {\includegraphics[width=3.4in]{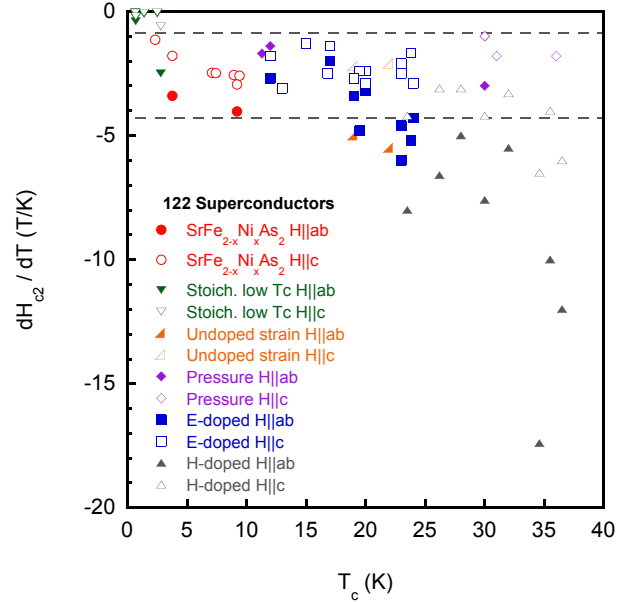}}
    \caption{(Color online) Upper critical field slope $\frac{\partial{H_{c2}}}{\partial{T}}$ of SrFe$_{2-x}$Ni$_{x}$As$_{2}$ compared to values from the literature. Solid symbols denote $H$ in-plane, while open symbols denote $H \parallel c$. For a large class of 122 superconductors, values of $\frac{\partial{H_{c2}}}{\partial{T}}$ are comparable despite a wide range of $T_c$ values.  Data for SrFe$_{2-x}$Ni$_{x}$As$_{2}$ are from this work. Stoichiometric low $T_c$ materials include SrNi$_2$P$_2$,\cite{Ronning09} SrNi$_2$As$_2$,\cite{Bauer08} BaNi$_2$As$_2$,\cite{Ronning08} KFe$_2$As$_2$,\cite{Terashima09K} BaNi$_2$As$_2$,\cite{Kurita09} and BaNi$_2$P$_2$.\cite{Tomioka09} The other superconductors are discussed in the text.  Data for ``undoped strain'' are from Refs.~\onlinecite{Saha09PRL,Kim09Influx}, ``pressure'' from Refs.~\onlinecite{Colombier09,Terashima09,Kotegawa09,Torikachvili08,Torikachvili09}, ``electron-doped'' from Refs.~\onlinecite{Baily09,Choi09,Kim09,Ni08, Tanatar09, Nakajima09,Kumar09,Kumar09Ni}, and ``hole-doped'' from Refs.~\onlinecite{Chen08,Bukowski09,Welp09,Kim09Ba,Altarawneh08,Wang08,Ni08BaK,Yuan09}.}
    \label{Hc2lit}
\end{figure}

The superconducting upper critical field $H_{c2}$, with $H$ applied parallel to the $c$-axis, is shown in Fig.~\ref{Hc2} for concentrations across the superconducting range.   Here, $H_{c2}$ is defined where the $R(T)$ data have half the normal state value (0.5 in Fig.~\ref{Hc2sweeps}). With the exception of some curvature at low field, which is also seen in resistively determined values for other 122 superconductors,\cite{Ni08,Terashima09} the $H_{c2}(T)$ curves are strikingly linear.  The slope $\frac{\partial{H_{c2}}}{\partial{T}}$ ranges from -1.8~T/K for $x=0.12$ to a maximum value of -2.9~T/K for $x=0.16$ at optimal doping, to -1.1~T/K for $x=0.22$.  These slopes are comparable to those reported for BaFe$_2$As$_2$ and SrFe$_2$As$_2$ under pressure,\cite{Colombier09} EuFe$_2$As$_2$ under pressure,\cite{Terashima09} Co-substituted SrFe$_2$As$_2$ thin films,\cite{Baily09,Choi09} SrFe$_{1.6}$Co$_{0.4}$As$_2$,\cite{Kim09} Ba(Fe$_{1-x}$Co$_x$)$_2$As$_2$,\cite{Ni08, Tanatar09, Nakajima09} CaFe$_{1.94}$Co$_{0.06}$As$_2$,\cite{Kumar09} BaFe$_{1.91}$Ni$_{0.09}$As$_2$,\cite{Sun09} CaFe$_{1.94}$Ni$_{0.06}$As$_2$,\cite{Kumar09Ni} and ambient-pressure undoped strain-induced SrFe$_2$As$_2$ \cite{Saha09PRL} and BaFe$_2$As$_2$,\cite{Kim09Influx} as illustrated in Fig.~\ref{Hc2lit}. This agreement is rather remarkable, given that these superconductors have values of $T_c$ ranging from 20~K to more than 30~K, while the maximum value of $T_c$ for SrFe$_{2-x}$Ni$_{x}$As$_{2}$ is less than 10~K. Moreover, the response of the superconducting state to applied $H$ seems insensitive to whether superconductivity has been stabilized by transition metal substitution, applied pressure, or strain,\cite{Saha09PRL,Kim09Influx} in the case of the undoped parent compounds.  In contrast, hole-doped SrFe$_2$As$_2$ and BaFe$_2$As$_2$ feature larger values of $\frac{\partial{H_{c2}}}{\partial{T}}$.\cite{Chen08,Bukowski09,Welp09,Kim09Ba,Altarawneh08,Wang08,Ni08BaK,Yuan09}

The inset of Fig.~\ref{Hc2} shows the $H_{c2}(T)$ curves scaled with respect to $T_{c0}$, the value of $T_c$ at $H=0$, and the reduced field $H^*=H_{c2}(-\frac{\partial{H_{c2}}}{\partial{T}}T_{c0})^{-1}$, which is defined such that the slope $\frac{\partial{H^*}}{\partial{(T_c/T_{c0})}} = -1$. As shown in the inset of Fig.~\ref{Hc2}, the scaling procedure collapses all $H_{c2}(T)$ data onto one curve, underscoring the already-noted similarity in $H$-dependence for all superconducting concentrations of SrFe$_{2-x}$Ni$_{x}$As$_{2}$.  However, this plot also makes it clear that the Werthamer-Helfland-Hohenberg (WHH) estimate that $H_{c2}(0) \approx -0.7(T_{c0})\frac{\partial{H_{c2}}}{\partial{T}}$ is too low, because in SrFe$_{2-x}$Ni$_{x}$As$_{2}$, $T_c / T_{c0}=0.3$ at $H^*=0.7$, where the WHH model predicts $T_c$  should be 0.\cite{Werthamer66}  At this low reduced temperature, the scaled upper critical curve is still linear, and shows no signs of saturation.  In the presence of spin-orbit scattering and significant spin susceptibility, the value of $H_{c2}(0)$ would typically be suppressed with respect to the bare WHH estimate. In addition, the weak-coupling estimate of the paramagnetic critical field $H_P \approx 1.84T_c$ yields values comparable to those determined by the WHH formula.  In other words, both conventional estimates of the paramagnetic and orbital limiting fields underestimate the actual $H_{c2}(0)$. This characteristic has been noted in other iron pnictide superconductors and is discussed in a recent review.\cite{Putti09}

An extrapolation of the linear $H_{c2}(T)$ slope to $T=0$~K yields a nominal limiting value of 25~T for $x=0.16$. Actually, this extrapolation may  be quite accurate, as an almost linear $H_{c2}(T)$ is observed for $H \parallel c$ in Co-substituted SrFe$_2$As$_2$ thin films.\cite{Baily09} Superconducting coherence lengths $\xi = (\Phi_0 /2\pi H_{c2})^{1/2}$, where $\Phi_0$ is the flux quantum, range from about 10~nm to 3.5~nm for the range of $H_{c2}(0)$ values observed in SrFe$_{2-x}$Ni$_{x}$As$_{2}$.  For comparison, the value of $\xi = 3$~nm was determined for BaFe$_{1.8}$Co$_{0.2}$As$_{2}$ via scanning tunneling microscopy.\cite{Yin09} Note that the low-$H$ tails of the $H_{c2}(T)$ curves, which  are  rather large for $x=0.14$ and 0.15, have been ignored in the preceding analysis.  Instead, the $H \rightarrow 0$ limiting slope has been fit above this region of large curvature, because estimating $H_{c2}(0)$ using the low-$H$ portion of the $H_{c2}(T)$ curves leads to an erroneous dramatic underestimate.

\begin{figure}%[hp]
    {\includegraphics[width=3.4in]{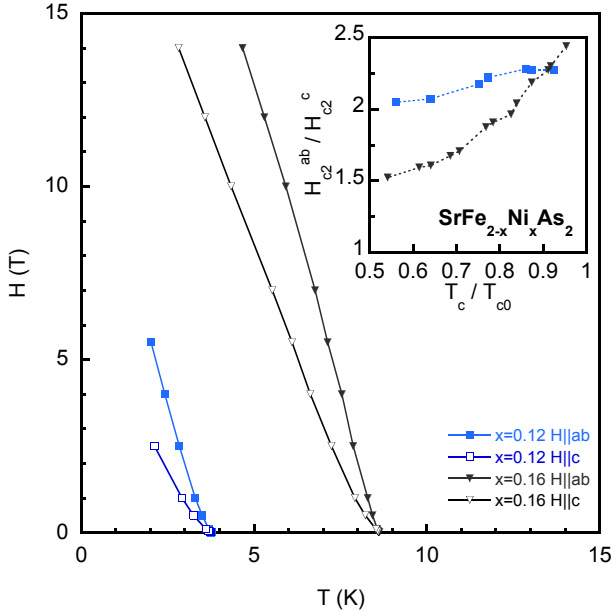}}
    \caption{(Color online) Upper critical field $H_{c2}$ of SrFe$_{2-x}$Ni$_{x}$As$_{2}$, determined via resistivity, for different field orientations. Inset: Anisotropy of $H_{c2}$, which shows stronger temperature dependence at optimal doping.}
    \label{anisotropy}
\end{figure}

The $H_{c2}(T)$ curves were also studied with $H$ applied perpendicular to the $c$-axis.  In Fig.~\ref{anisotropy}, a field-direction comparison is made for $x=0.12$ and 0.16.  As with all iron-based 122 superconductors, $H$ applied perpendicular to the $c$-axis suppresses $T_c$ less quickly in SrFe$_{2-x}$Ni$_{x}$As$_{2}$. Especially visible for $x=0.16$, there is increased curvature in $H_{c2}(T)$ relative to the $H \parallel c$ case, which is again consistent with measurements on Co-substituted SrFe$_2$As$_2$ thin films,\cite{Baily09} although due to the limited field and temperature range of the current measurements, it is unclear whether the $H_{c2}(T)$ curves extrapolate to the same value of $H$.  The anisotropy $H_{c2}^{ab} / H_{c2}^c$ is plotted in the inset of Fig.~\ref{anisotropy}. The anisotropy in $x=0.12$ is roughly constant at a value of about 2.2, whereas the anisotropy in $x=0.16$ varies from approximately 2.5 near $T_c$ to 1.5 at $0.5T_c$.  This range of anisotropy values is consistent with both electron-\cite{Ni08,Tanatar09} and hole-doped 122 materials.\cite{Bukowski09,Welp09,Yuan09} Considering the range of values of $T_c$ exhibited by these superconductors, the similarity is again noteworthy.

\begin{figure}%[hp]
    {\includegraphics[width=3.4in]{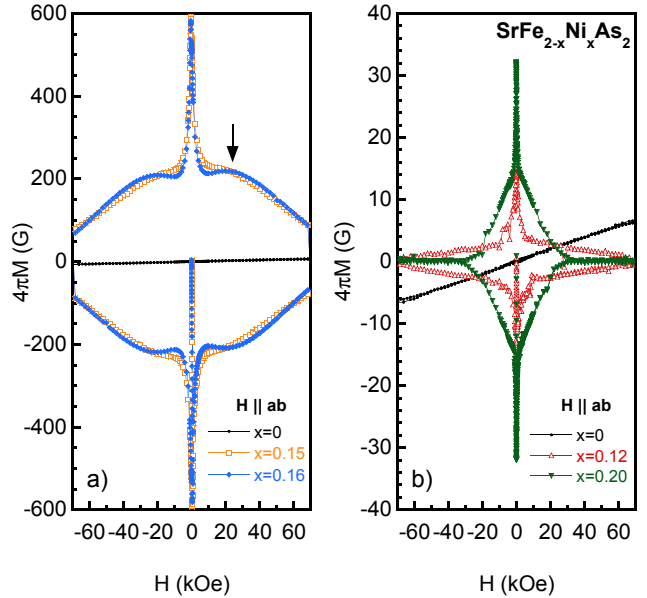}}
    \caption{(Color online) Magnetization of SrFe$_{2-x}$Ni$_{x}$As$_{2}$ as a function of applied field at 1.8~K, with $H$ applied in-plane. a) Hysteresis loops for optimally doped samples have a large area and values of $H_{c2}$ exceeding 7~T. The $x=0.16$ data exhibit a secondary peak, identified by the arrow.  b) For under- and over-doped samples, the loops are smaller and critical fields are lower. $M(H)$ data for the non-superconducting parent compound are shown for comparison. Note the different vertical scales. 10~kOe$=$1~T. }
    \label{MHloops}
\end{figure}

\section{Magnetic properties}
The field-dependence of the superconducting state was also investigated via measurements of the magnetization $M(T,H)$. The low-$H$ $M(T)$ data show that SrFe$_{2-x}$Ni$_{x}$As$_{2}$ exhibits a full Meissner effect near optimal doping, for $0.14 \leq x \leq 0.16$.\cite{Saha09}  In order to further compare the magnetic response of the superconducting state, the $H$-dependence at 1.8~K was studied for $x$ across the superconducting dome. As Fig.~\ref{MHloops} shows, there is a clear difference between optimally doped samples, and the under- and over-doped samples.  Note that the $x=0.12$ and  $x=0.20$ data have had a normal-state contribution subtracted because the superconductivity does not fully occupy the bulk. Aside from expected differences in $H_{c2}$ and the lower critical field $H_{c1}$, the area of the $M(H)$ loops is much larger near optimal doping.  The $x=0.16$ sample exhibits a secondary peak, as has been observed in other electron-doped 122 compounds, although the local minima at low $H$ can be much deeper in other systems.\cite{Prozorov08,Sun09} In contrast, this phenomenon is not observed in the $x=0.15$ sample. Such fishtail structures in $M(H)$ hysteresis loops are thought to arise from vortex pinning effects, which raises an interesting question about whether the appearance of this behavior for $x=0.16$ may be associated with the disappearance of magnetic order near $x=0.15$ (Fig.~\ref{phsdgm}). However, a more systematic study is required to rule out simple disorder effects or sample dependence.

Turning to the $M(H)$ data for $x=0.12$, it is evident that the apparent value $H_{c2}$ exceeds 7~T, as the $M(H)$ loop has nonzero area up to 7~T, whereas the superconducting transition appears to be already suppressed by this field in $\rho(T)$ data (Figs.~\ref{Hc2sweeps} and \ref{Hc2}).  This effect may be the result of macroscopic phase separation between magnetic and superconducting regions in the sample, where there is no superconductive path between isolated superconducting sections.  Alternatively, the nonzero area in $M(H)$ may be due to a small, uncorrected-for ferromagnetic contribution, similar to that observed in undoped SrFe$_2$As$_2$.\cite{Saha09PRL}  In contrast, the $x=0.20$ sample exhibits the opposite: a much lower $H_{c2}$ in $M(H)$ than in resistance data, which can be ascribed to the persistence of tiny connected regions of supercurrent that offer ineffective diamagnetic screening.  In addition, the $M(H)$ data for $x=0.18$, which are not shown, exhibit inhomogeneous superconductivity, despite being chemically homogeneous as determined by energy dispersive x-ray spectroscopy.\cite{Saha09}  There exists a large variation in $T_c$ and volume fraction between samples for $x \approx 0.12$ and $0.18 \leq x \leq 0.2$, suggesting that the superconducting region of the phase diagram is not really a rounded dome, but rather that the superconductivity onset and offset as a function of $x$ are sharp.  Again, local probes such as $\mu$SR or NMR are required to confirm this scenario.

\begin{figure}%[hp]
    {\includegraphics[width=3.4in]{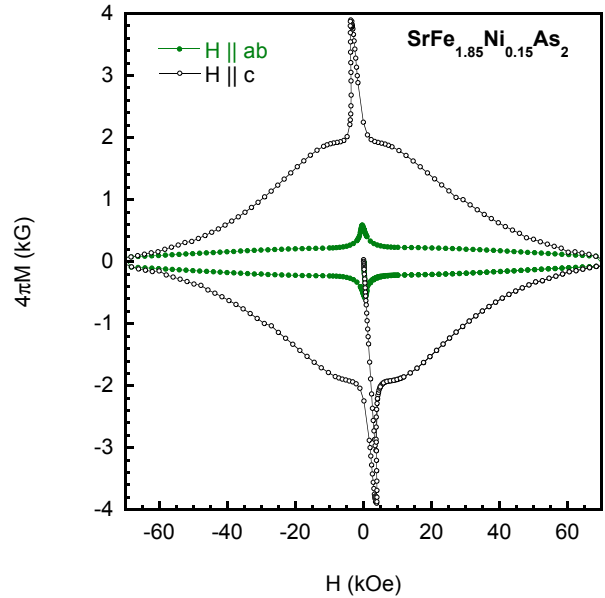}}
    \caption{(Color online) Direction dependence of the magnetization of SrFe$_{1.85}$Ni$_{0.15}$As$_{2}$ at 1.8~K. Field applied in-plane evokes a much larger magnetic response than out-of-plane, and the lower critical field $H_{c1}$ is larger for $H \parallel c$. Demagnetization effects have been corrected for. 10~kOe$=$1~T.}
    \label{MHdircomp}
\end{figure}

It is instructive to check whether the anisotropy of the superconducting state evident from transport measurements is also seen in the $M(H)$ data.  Figure~\ref{MHdircomp} confirms that there is sizable magnetic anisotropy for optimally-doped $x=0.15$.  (Demagnetization effects have  been corrected for.) With $H \parallel c$, the magnetization in the superconducting state is almost an order of magnitude larger than with $H$ in-plane. This behavior is consistent with that observed in BaFe$_{1.8}$Co$_{0.2}$As$_{2}$.\cite{Sefat08} Concomitantly, the value of $H_{c1}$, as estimated from the local minimum in the virgin $M(H)$ curve, is larger for $H \parallel c$, with  $H_{c1}^c = 0.36$~T, than for $H \perp c$, with $H_{c1}^{ab} =0.08$~T. The initial slope of the virgin curve is consistent between the two $H$ orientations and indicates full diamagnetic screening.

\begin{figure}%[hp]
    {\includegraphics[width=3.4in]{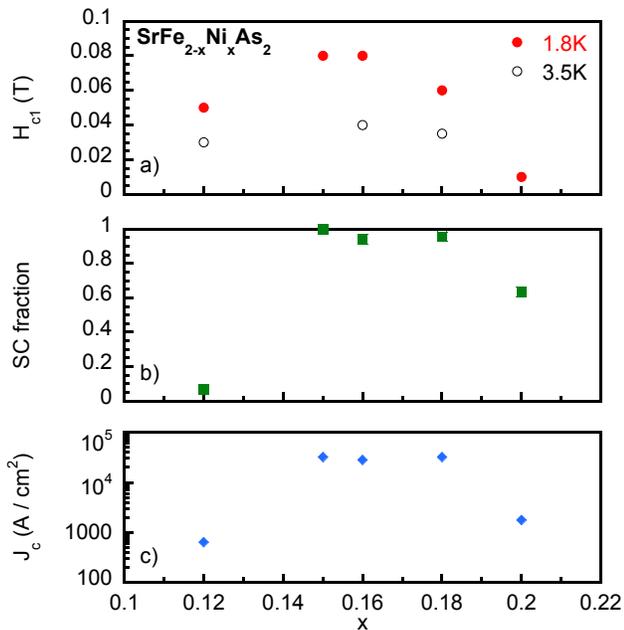}}
    \caption{(Color online) Concentration dependence of the low-temperature properties of the superconducting state of SrFe$_{2-x}$Ni$_{x}$As$_{2}$. a) The lower critical field $H_{c1}$, identified with the local minimum in the $M(H)$ virgin curve. b) The superconducting volume fraction calculated from the initial slope of the virgin $M(H)$ curves at 1.8~K.  c) Superconducting critical current $J_c$, calculated using the Bean model.}
    \label{MHparams}
\end{figure}

A summary of quantities derived from the low-$T$ $M(H)$ data for $H \perp c$ is presented in Fig.~\ref{MHparams}. The values of $H_{c1}$ were determined from the local minimum in the virgin $M(H)$ data.  These values are about a factor of 4 larger than the value of $H$ at which $M(H)$ starts to deviate from linearity.  This alternative criterion for defining  $H_{c1}$ is more difficult to apply precisely because of curvature in $M(H)$ at fields below the local minimum. Estimates of the superconducting volume fraction are based on the initial slope of the virgin $M(H)$ curves and are consistent with values determined from low-$H$ $M(T)$ data.\cite{Saha09} Magnitudes of the critical current $J_c$ have been estimated using the Bean model.  For the optimally-doped samples, $J_c = 3\times 10^4$~A~cm$^{-2}$, with $H \parallel ab$, while for $H \parallel c$,  $J_c = 2\times 10^4$~A~cm$^{-2}$. These values are about one order of magnitude lower than values quoted for Ba(Fe,Co)$_2$As$_2$,\cite{Sefat08,Prozorov08,Tanatar09,Nakajima09,Sun09} BaFe$_{1.91}$Ni$_{0.09}$As$_{2}$,\cite{Sun09} and BaFe$_2$(As$_{0.68}$P$_{0.32}$)$_{2}$.\cite{Chong09} The hole-doped 122 superconductors exhibit significantly higher values of $J_c \approx 10^6$~A~cm$^{-2}$.\cite{Bukowski09,Kim09Ba}  Combining the values of $H_{c1}$ and $H_{c2}$, it is possible to roughly estimate the magnitude of the penetration depth $\lambda =  \frac{H_{c2} \xi}{\surd2 H_c}$, where the thermodynamic critical field $H_c \approx (H_{c1}H_{c2})^{\frac{1}{2}}$. For both $x=0.12$ and $x=0.16$ with $H \parallel ab$, $\lambda^{ab} = 50$~nm, while for $x=0.15$ with $H \parallel c$, $\lambda^{c} = 20$~nm.  In contrast, in SrFe$_{1.75}$Co$_{0.25}$As$_2$, values of $\lambda^{ab} = 315$~nm and $\lambda^{c} = 870$~nm were determined by $\mu$SR measurements.\cite{Khasanov09}

\section{Discussion}
The present study has uncovered several interesting properties of the SrFe$_{2-x}$Ni$_{x}$As$_{2}$ system.  First, neither the $R_H(T)$ or $S(T)$ data reflect an obvious increase in carriers, although both measurements appear to be dominated by negatively-charged carriers.  Second, despite the relatively lower values of $T_c$, there is nothing strikingly different about the superconducting state in SrFe$_{2-x}$Ni$_{x}$As$_{2}$, at least in the bulk properties that have been probed in this study.  However, it is intriguing that $\frac{\partial{H_{c2}}}{\partial{T}}$ values are comparable to those in electron-doped, pressure-induced, and strain-induced 122 superconductors, despite the variation in $T_c$ (Fig.~\ref{Hc2lit}).

Recent angle-resolved photoemission measurements have revealed that in both K- and Co-substituted BaFe$_2$As$_2$, superconductivity requires the presence of both electron and hole pockets, and that the nesting conditions differ between hole- and electron-substituted materials.\cite{Brouet09,Ding08} This difference may explain the variation in $T_c$ values, and perhaps even $H_{c2}$ slopes, between the cases of hole and electron doping in the 122 materials. The multiband nature of the superconductivity can also explain the inapplicability of standard estimates of spin- and orbital-limiting critical fields.\cite{Putti09} It might further be expected that the disappearance of superconductivity would coincide with a change in Fermi surface. Indeed, it is argued  that the temperature dependence of the Hall effect disappears at the end of the superconducting dome in Ba(Fe$_{1-x}$Co$_x$)$_2$As$_2$.\cite{Fang09}  However, in the case of SrFe$_{2-x}$Ni$_{x}$As$_{2}$, such an effect is not observed, as the values of $R_H$ look to be independent of $x$ to within experimental error. Moreover, there is no dramatic change in $S(T)$ as a function of $x$ either, which suggests that variations in the Fermi surface between superconducting and normal states in SrFe$_{2-x}$Ni$_{x}$As$_{2}$ are more subtle.

This absence of obvious changes in the Fermi surface is possibly related to the relatively narrow superconducting dome and the low values of $T_c$. The Fermi surface topology required for superconductivity, e.g. for effective hopping in an $s_{\pm}$ model,\cite{Chubukov09}  may be less well realized  in SrFe$_{2-x}$Ni$_{x}$As$_{2}$ compared to other 122 superconductors, over a smaller range of $x$, leading to a smaller number of superconducting carriers and lower $T_c$.  This simple explanation does not, however, account for the similarity in $\frac{\partial{H_{c2}}}{\partial{T}}$ values. The apparent similarities between the superconducting states in different 122 materials must also be interpreted cautiously.  For example, in both SrFe$_{2-x}$Ni$_{x}$As$_{2}$ and pressure-induced 122 superconductors,  full diamagnetic screening is observed over only part of the entire superconducting range.\cite{Alireza09,Colombier09}  However, this is not true for all electron-doped 122 compounds, because in Ba(Fe$_{1-x}$Co$_x$)$_2$As$_2$, full screening is observed for all superconducting concentrations.\cite{Gordon09} Currently, it is not clear whether such differences are intrinsic, or sample- and measurement-dependent.  Experimental and theoretical investigation of the differences between the Fermi surfaces in Co- and Ni-substituted SrFe$_2$As$_2$ may help shed light on this issue.

To summarize, Hall effect and thermoelectric power measurements on SrFe$_{2-x}$Ni$_{x}$As$_{2}$  are consistent with a dominant electron-like response, as is seen in other transition metal substitution studies of 122 superconductors.  The slope of the upper critical field $\frac{\partial{H_{c2}}}{\partial{T}}$ is linear over a wide temperature range and appears to defy conventional estimates of paramagnetic and orbital limits. Despite lower values of $T_c$ in this Ni-substitution series, values of $\frac{\partial{H_{c2}}}{\partial{T}}$ are comparable to other electron-doped 122 superconductors with higher $T_c$. For samples exhibiting maximum $T_c \approx 9$~K, full volume fraction superconductivity is observed, while samples exhibiting lower $T_c$ seem to exhibit inhomogeneous superconductivity, with disconnected superconducting regions on the underdoped side and  tiny connected regions of superconductivity on the overdoped side. Further investigation of possible magnetic/superconducting phase separation and details of the Fermi surface evolution are called for.

\begin{acknowledgments}

XHZ and RLG are supported by the NSF under Grant No. DMR-0653535. NPB is supported by CNAM.
\end{acknowledgments}


\begin{thebibliography}{99}

\bibitem{Parker09} D. Parker, M. G. Vavilov, A. V. Chubukov, and I. I. Mazin, Phys. Rev. B 80, 100508(R) (2009).

\bibitem{Chubukov09} A. V. Chubukov, M. G. Vavilov, and A. B. Vorontsov, Phys. Rev. B 80, 140515(R) (2009).

\bibitem{Butch08} N. P. Butch, M. C. de Andrade, and M. B. Maple, Amer. J. Phys. 76, 106 (2008).

\bibitem{Saha09} S. R. Saha, N. P. Butch, K. Kirshenbaum, and Johnpierre Paglione, Phys. Rev. B 79, 224519 (2009).

\bibitem{Han09} F. Han, X. Zhu, P. Cheng, G. Mu, Y. Jia, L. Fang, Y. Wang, H. Luo, B. Zeng, B. Shen, L. Shan, C. Ren, and H.-H. Wen, Phys. Rev. B 80, 024506 (2009).

\bibitem{Schnelle09} W. Schnelle, A. Leithe-Jasper, R. Gumeniuk, U. Burkhardt, D. Kasinathan, and H. Rosner, Phys. Rev. B 79, 214516 (2009).

\bibitem{Ni09} N. Ni, A. Thaler, A. Kracher, J. Q. Yan, S. L. Bud'ko, and P. C. Canfield, Phys. Rev. B 80, 024511 (2009).

\bibitem{Budko09} S. L. Bud'ko, N. Ni, and P. C. Canfield, Phys. Rev. B 79, 220516(R) (2009)

\bibitem{Rullier-Albenque09} F. Rullier-Albenque, D. Colson, A. Forget, and H. Alloul, Phys. Rev. Lett. 103, 057001 (2009).

\bibitem{Fang09} L. Fang, H. Luo, P. Cheng, Z. Wang, Y. Jia, G. Mu, B. Shen, I. I. Mazin, L. Shan, C. Ren, and H.-H. Wen, Phys. Rev. B 80, 140508(R) (2009).

\bibitem{Yan08} J.-Q. Yan, A. Kreyssig, S. Nandi, N. Ni, S. L. Bud'ko, A. Kracher, R. J. McQueeney, R. W. McCallum, T. A. Lograsso, A. I. Goldman, and P. C. Canfield, Phys. Rev. B 78, 024516 (2008).

\bibitem{Mun09} E. D. Mun, S. L. Bud'ko, N. Ni, A. N. Thaler, and P. C. Canfield, Phys. Rev. B 80, 054517 (2009).

\bibitem{Brouet09} V. Brouet, M. Marsi, B. Mansart, A. Nicolaou, A. Taleb-Ibrahimi, P. Le F\`{e}vre, F. Bertran, F. Rullier-Albenque, A. Forget, and D. Colson, Phys. Rev. B 80, 165115 (2009).

\bibitem{Liu09} C. Liu, T. Kondo, R. M. Fernandes, A. D. Palczewski, E. D. Mun, N. Ni, A. N. Thaler, A. Bostwick, E. Rotenberg, J. Schmalian, S. L. Bud'ko, P. C. Canfield, and A. Kaminski, arXiv:0910.1799 (unpublished).

\bibitem{Matusiak09} M. Matusiak, Z. Bukowski, and J. Karpinski, Phys. Rev. B 81, 020510 (2010).

\bibitem{Park09}  J. T. Park, D. S. Inosov, Ch. Niedermayer, G. L. Sun, D. Haug, N. B. Christensen, R. Dinnebier, A.V. Boris, A. J. Drew, L. Schulz, T. Shapoval, U. Wolff, V. Neu, Xiaoping Yang, C. T. Lin, B. Keimer, and V. Hinkov, Phys. Rev. Lett. 102, 117006 (2009).

\bibitem{Laplace09} Y. Laplace, J. Bobroff, F. Rullier-Albenque, D. Colson, and A. Forget, Phys. Rev. B 80, 140501(R) (2009).

\bibitem{Pratt09}  D. K. Pratt, W. Tian, A. Kreyssig, J. L. Zarestky, S. Nandi, N. Ni, S. L. Bud'ko, P. C. Canfield, A. I. Goldman, and R. J. McQueeney, Phys. Rev. Lett 103, 087001 (2009).

\bibitem{Ni08}  N. Ni, M. E. Tillman, J.-Q. Yan, A. Kracher, S. T. Hannahs, S. L. Bud'ko, and P. C. Canfield, Phys. Rev. B 78, 214515 (2008).

\bibitem{Terashima09} T. Terashima, M. Kimata, H. Satsukawa, A. Harada, K. Hazama,S. Uji, H. S. Suzuki, T. Matsumoto, and K. Murata, J. Phys. Soc. Jpn. 78, 083701 (2009).

\bibitem{Colombier09} E. Colombier, S. L. Bud'ko, N. Ni, and P. C. Canfield, Phys. Rev. B 79, 224518 (2009).

\bibitem{Baily09} S. A. Baily, Y. Kohama, H. Hiramatsu, B. Maiorov, F. F. Balakirev, M. Hirano, and H. Hosono, Phys. Rev. Lett. 102, 117004 (2009).

\bibitem{Choi09} E.-M. Choi, S.-G. Jung, N. H. Lee, Y.-S. Kwon, W. N. Kang, D. H. Kim, M.-H. Jung, S.-I. Lee, and L. Sun, Appl. Phys. Lett. 95, 062507 (2009).

\bibitem{Kim09}  J. S. Kim, S. Khim, L. Yan, N. Manivannan, Y. Liu, I. Kim, G. R. Stewart, and K. H. Kim, J. Phys.: Condens. Matter 21, 102203 (2009).

\bibitem{Tanatar09}  M. A. Tanatar, N. Ni, C. Martin, R. T. Gordon, H. Kim, V. G. Kogan, G. D. Samolyuk, S. L. Bud'ko, P. C. Canfield, and R. Prozorov, Phys. Rev. B 79, 094507 (2009).

\bibitem{Nakajima09} Y. Nakajima, T. Taen, and T. Tamegai, J. Phys. Soc. Jpn. 78, 023702 (2009).

\bibitem{Kumar09} N. Kumar, R. Nagalakshmi, R. Kulkarni, P. L. Paulose, A. K. Nigam, S. K. Dhar, and A. Thamizhavel, Phys. Rev. B 79, 012504 (2009).

\bibitem{Sun09} D. L. Sun, Y. Liu, and C. T. Lin, Phys. Rev. B 80, 144515 (2009).

\bibitem{Kumar09Ni} N. Kumar, S. Chi, Y. Chen, K. G. Rana, A. K. Nigam, A. Thamizhavel, W. Ratcliff, S. K. Dhar, and J. W. Lynn, Phys. Rev. B 80, 144524 (2009).

\bibitem{Saha09PRL} S. R. Saha, N. P. Butch, K. Kirshenbaum, J. Paglione, and P. Y. Zavalij, Phys. Rev. Lett. 103, 037005 (2009).

\bibitem{Kim09Influx} J. S. Kim, T. D. Blasius, E. G. Kim, and G. R. Stewart,  J. Phys.: Condens. Matter 21, 342201 (2009).

\bibitem{Chen08}  G. F. Chen, Z. Li, J. Dong, G. Li, W. Z. Hu, X. D. Zhang, X. H. Song, P. Zheng, N. L. Wang, and J. L. Luo, Phys. Rev. B 78, 224512 (2008).

\bibitem{Bukowski09}  Z. Bukowski, S. Weyeneth, R. Puzniak, P. Moll, S. Katrych, N. D. Zhigadlo, J. Karpinski, H. Keller, and B. Batlogg, Phys. Rev. B 79, 104521 (2009).

\bibitem{Welp09} U. Welp, R. Xie, A. E. Koshelev, W. K. Kwok, H. Q. Luo, Z. S. Wang, G. Mu, and H. H. Wen, Phys. Rev. B 79, 094505 (2009).

\bibitem{Kim09Ba} H.-J. Kim, Y. Liu, Y. S. Oh, S. Khim, I. Kim, G. R. Stewart, and K. H. Kim, Phys. Rev. B 79, 014514 (2009).

\bibitem{Altarawneh08}  M. M. Altarawneh, K. Collar, C. H. Mielke, N. Ni, S. L. Bud'ko, and P. C. Canfield, Phys. Rev. B 78, 220505(R) (2008).

\bibitem{Wang08}  Z.-S. Wang, H.-Q. Luo, C. Ren, and H.-H. Wen, Phys. Rev. B 78, 140501(R) (2008).

\bibitem{Ni08BaK}  N. Ni, S. L. Bud'ko, A. Kreyssig, S. Nandi, G. E. Rustan, A. I. Goldman, S. Gupta, J. D. Corbett, A. Kracher, and P. C. Canfield, Phys. Rev. B 78, 014507 (2008).

\bibitem{Yuan09} H. Q. Yuan, J. Singleton, F. F. Balakirev, S. A. Baily, G. F. Chen, J. L. Luo, and  N. L. Wang, Nature (London) 457, 565 (2009).

\bibitem{Ronning09}  F. Ronning, E. D. Bauer, T. Park, S.-H. Baek, H. Sakai, and J. D. Thompson, Phys. Rev. B 79, 134507 (2009).

\bibitem{Bauer08} E. D. Bauer, F. Ronning, B. L. Scott, and J. D. Thompson, Phys. Rev. B 78, 172504 (2008).

\bibitem{Ronning08} F. Ronning, N. Kurita, E. D. Bauer, B. L. Scott, T. Park, T. Klimczuk, R. Movshovich, and J. D. Thompson, J. Phys.: Condens. Matter 20, 342203 (2008).

\bibitem{Terashima09K} T. Terashima, M. Kimata, H. Satsukawa, A. Harada, K. Hazama, S. Uji, H. Harima, G.-F. Chen, J.-L. Luo, and N.-L. Wang, J. Phys. Soc. Jpn. 78, 063702 (2009).

\bibitem{Kurita09}  N. Kurita, F. Ronning, Y. Tokiwa, E. D. Bauer, A. Subedi, D. J. Singh, J. D. Thompson, and R. Movshovich, Phys. Rev. Lett. 102, 147004 (2009).

\bibitem{Tomioka09}  Y. Tomioka, S. Ishida, M. Nakajima, T. Ito, H. Kito, A. Iyo, H. Eisaki, and S. Uchida. Phys. Rev. B 79, 132506 (2009).

\bibitem{Kotegawa09}  H. Kotegawa, H. Sugawara, and H. Tou, J. Phys. Soc. Jpn. 78, 013709 (2009).

\bibitem{Torikachvili08}  M. S. Torikachvili, S. L. Bud'ko, N. Ni, and P. C. Canfield, Phys. Rev. Lett. 101, 057006 (2008).

\bibitem{Torikachvili09} M. S. Torikachvili, S. L. Bud'ko, N. Ni, P. C. Canfield, and  S. T. Hannahs, Phys. Rev. B 80, 014521 (2009).

\bibitem{Werthamer66} N. R. Werthamer, E. Helfland, and P. C. Hohenberg, Phys. Rev. 147, 295 (1966).

\bibitem{Putti09} M. Putti, I. Pallecchi, E. Bellingeri, M. Tropeano, C. Ferdeghini, A. Palenzona, C. Tarantini, A. Yamamoto, J. Jiang, J. Jaroszynski, F. Kametani, D. Abraimov, A. Polyanskii, J. D. Weiss, E. E. Hellstrom, A. Gurevich, D. C. Larbalestier, R. Jin, B. C. Sales, A. S. Sefat, M. A. McGuire, D. Mandrus, P. Cheng, Y. Jia, H. H. Wen, S. Lee, and C. B. Eom, arXiv:0910.1297 (unpublished).

\bibitem{Yin09} Y. Yin, M. Zech, T. L. Williams, X. F. Wang, G. Wu, X. H. Chen, and J. E. Hoffman, Phys. Rev. Lett. 102, 097002 (2009).

\bibitem{Prozorov08}  R. Prozorov, N. Ni, M. A. Tanatar, V. G. Kogan, R. T. Gordon, C. Martin, E. C. Blomberg, P. Prommapan, J. Q. Yan, S. L. Bud'ko, and P. C. Canfield, Phys. Rev. B 78, 224506 (2008).

\bibitem{Sefat08} A. S. Sefat, R. Jin, M. A. McGuire, B. C. Sales, D. J. Singh, and D. Mandrus, Phys. Rev. Lett. 101, 117004 (2008).

\bibitem{Chong09} S. V. Chong, S. Hashimoto, and K. Kadowaki, arXiv:0908.3284 (unpublished).

\bibitem{Khasanov09} R. Khasanov, A. Maisuradze, H. Maeter, A. Kwadrin, H. Luetkens, A. Amato, W. Schnelle, H. Rosner, A. Leithe-Jasper, and H.-H. Klauss, Phys. Rev. Lett. 103, 067010 (2009).

\bibitem{Ding08} H. Ding, P. Richard, K. Nakayama, K. Sugawara, T. Arakane, Y. Sekiba, A. Takayama, S. Souma, T. Sato, T. Takahashi, Z. Wang, X. Dai, Z. Fang, G. F. Chen, J. L. Luo, and N. L.Wang, EPL 83, 47001 (2008).

\bibitem{Alireza09} P. L. Alireza, Y. T. C. Ko, J. Gillett, C. M. Petrone, J. M. Cole, G. G. Lonzarich, and S. E. Sebastian, J. Phys.: Condens. Matter 21, 012208 (2009).

\bibitem{Gordon09}  R. T. Gordon, C. Martin, H. Kim, N. Ni, M. A. Tanatar, J. Schmalian, I. I. Mazin, S. L. Bud'ko, P. C. Canfield, and R. Prozorov, Phys. Rev. B 79, 100506(R) (2009).

\end{thebibliography}
\end{document}